\documentstyle[12pt]{article}
\begin{document}

\def\ds{\displaystyle}

\sloppy
\title{\bf Quantum Electronic Devices\\
       Based  on Metal-Dielectric Transition
        in  Low-Dimensional Quantum  Structures}
\author{{\bf I. Antoniou}\\ Solvay Institute, Brussels, Belgium\\
       {\bf B. Pavlov},\thanks{Actual address: Department  of  Mathematics, the
University of
Auckland, New  Zealand.}
{\bf  A. Yafyasov}\\
Institute for Physics, St. Petersburg University, Russia}
\date{}
\maketitle
\thispagestyle{empty}
{\bf 1. Introduction. }
Recently   P. Shor [1]  discovered  an  efficient
 quantum ``polynomial-time   algorithm" (length
$(\lg_{2} N)^k$ for  some  finite  $k$)
for    factoring  large  numbers  $N$
in the form  of  a  product  of prime  factors. This  event  makes
the implementation   of more  efficient  quantum  gates
(see [2]) an  actual
problem  of  quantum  engineering.
\par
 The  quality  of  the
implementation  of  the  gates  is  characterized   by  the  energy
 $E = \frac{1}{T_0}$  of  interaction  between  pair  of  qubits
 and  the  damping  rate  $\Gamma = \frac {1}{\tau}$ for  the
 coherence  of  any  one  qubit. The   ratio  of
  the  single  qubit coherence  decay  time $ \tau$ (``coherence  time")
 and  the time  $T_0$, required to  perform  the  coherent  change  of
 state  of  qbits due  to  the  interaction (``switching  time")
 should  be large  for  reversible  coherent  computations:
\[
\frac{\tau}{T_0}>>1.
\]
\par
The  quantum  coherence  has  been  observed  for  tunnel-coupled
 systems  of  several quantum   dots. The dominant  mechanism
destroying  coherence  at  low  temperatures  is the
 electron-electron  scattering, which  is  sensitive
 to  the  geometry  of  devices [3,4,5]. The  above  ratio  for
a  single  quantum  dot  is  approximately  $10^3$  at  Helium
temperatures  and could  be  improved  if  smaller   quantum
dots  could  be  constructed.
\par
The  superconducting  quantum interference  devices,
 constructed  as  a  system of  superconducting  rings  with
Josephson  contacts  between  them  are  dual  to  the  quantum
dots, trapping  magnetic  flux  instead  of  electrons,
however the
loss of  coherence  in  those  devices  is  not  properly
 understood  jet (see [6]).
\par
Cirac  and  Zoller [7]  proposed   recently   a laser-cooled
 array  of  Ions  in  the  linear  Pauly  trap
as  a  device  for   quantum
 computations. The  coherence  time for this device
could  be  several  seconds  or  even  longer, and
 the  switching  time  is  defined  by  the  optical
transitions  frequency.Thus  the  characteristic  ratio of
``coherence  time"  to ``switching  time"
should  be    large  at Helium  temperatures.
For  modern  discussion of other optical gates  see [8].
\par
It  was  shown (see for instance [9]), that  only
two types of  operations  -``gates"-  are  necessary  to  perform  any
 quantum  computation: the  single  bit  ``Rotate  Operator"  and
 two-bits  ``Controlled-Not Operator".
In this  paper  we  discuss  the  implementation  of
quantum  gates  ``Controlled-Not" in the form  of
a  quasi-molecular  chain  of  atoms  or  clusters,
controlled  by  some  sort  of  optical excitation.
 Our  analysis  shows,
that  for  proper  choice  of  materials the  working
temperatures  of  the  proposed  device
  might  be
essentially  higher, than  the  working  temperatures
of  devices  mentioned  above.
\par
 The  role  of  one  of  qubits in  our  device  is  played
  by the  inner atomic  electrons
which  are  transferred simultaneously by  optical  excitation
 in  all  atoms  of the  chain from  the  ground  state
into  the  excited  state, but  still  remain  inside atoms;
another  qubit  is  represented  by   the  asymptotic
states (scattering  states) of
the conductivity  electrons in  quantum  wires
attached  to  the  chain. In  dependence  of  the
occupation  of  inner   orbitals  the  transmission
coefficient  might  be  (close  to) zero
or, to  the  transmission  coefficient
through   semiinfinite periodic  lattice.
These  cases  correspond  to the
case  of  zero  conductance  of  the  device  or,
respectively, to  the  good  conductance  in  accordance
with  Landauer  formula [10,11].
The  corresponding  ratio  is  formed  by  the
life-time  of  the  excited  state  and  the
inverse  optical  frequency, and  could  vary  in a  wide
 range in  dependence  on  the  choice  of  materials.
 \par
 Strictly  speaking  the  scattering  states
of  conductivity  electrons   are
 not  representing  qubits  in
proper  sense, since  the  transmission
coefficient  through  the  chain  could  not  be
precisely   $0$  or  modulo  $1$. Nevertheless   we
can show, that the  proper  choice of  the  materials
 and the parameters  of  the  device
makes them  reasonable  close
to   $0$  or  to  the
transmission  coefficient $T_{\infty}$ for  semiinfinite
lattice    in  dependence  of
the  occupation  of  atomic  orbitals.
Due  to  the  unitarity of  the  complete
 scattering  matrix, the
difference  $ 1- |T_{\infty}|^2 $ is equal,
to  the  square  modulo  of  the  reflection coefficient
from  the  infinite  halflattice.
Formally  It  plays
the  role  of  the  relative measure  of  necessary
``garbage''  which  should  be  stored  and  used
when  performing  invertible  computations. But  in
our   case   the  reflection  coefficient   is  an
analytic  function  of   energy, hence   it   can
be  restored  from   the  orthogonality  condition,
in  unique  way, up  to  some    unimodular
Blashke-factor,  which  is  trivial, if  the
bound  states  with  negative  energies   are  absent.
Thus  the  nonzero  transmission  coefficient  contains
the   complete   information  of  the  incoming  signal,
and {\it nothing}  is  lost   in  form  of  reflected  wave,
hence  the  computation  act   is  reversible, though
some  part  of  the  signal  disappears  forever.
\par
The  {\it  irreversible}  loss  of  information
appeares  when
  measuring  the   result
of   scattering   by   some classical  device or  even
by  some  quantum  device   on   a  stochastic   background.
Then  the  part  of  information  contained  in  the
transmitted   signal   is  destroyed  and   the
computation   {\it becomes   irreversible}.
More  detailed  discussion  of  comparison  of
quantum  computations  and  classical  computations  on
stochastic (Markov)  background   is  contained
in  the  last  section  4.
\par
   We  are  discussing here  one-electron
 approximation, assuming, that  life-time
   of  the  inner   cluster  electrons
is  large  enough, hence  the
occupation  of  cluster  orbitals  is
  changing  by   resonance  laser  shining  only.
But  our  model  can  be  used
beyond  these  limits  as  well, if  we  consider
the  corresponding  two-electrons  problem, as
in [12]. It  will be  done  elsewhere.
\vskip0.5cm
{\bf 2. One-electron  scattering  by  single quantum  dot: Soluble  model
and  physical  reality}.
In  this  section
we  describe a  soluble  model  of  the  simplest
quantum ``Controlled-not"  quantum  gate
represented  in the form  a  quasi-molecular  cluster  attached
to  some  quantum  wire  by  tunnel contacts. This  model  can  be  applied
for  theoretical  analysis  of  constructions  of  nanoelectronic
devices  of  narrow-band  semiconductors.
\par
We  represent  the one-electron  Hamiltonian
of quantum  dot  attached  to  the  quantum
wire  as  a  result  of  perturbation
of  the  orthogonal  sum  of  ``internal"
and  ``external"  Hamiltonians.
The ``internal" Hamiltonian
of the electron in a cluster is  determined by the
occupation of the molecular
orbitals
$$
A_g \sim \alpha_e < \alpha_1 < \alpha_2 <\dots ;
\quad
A_g\phi_s^g =
\alpha_s\phi_s^g
$$
$$
A_e\sim
\alpha_g < \alpha_1 < \alpha_2 <\dots ;
\quad
A_e\phi_s^e =
\alpha_s\phi_s^e.
$$
Before  the  interaction  is  switched  on,
we  represent  the   internal
Hamiltonian  by  proper finite-dimensional
operators (diagonal  matrices)  in  each  ground
and  excited  states.
The   ``external"  Hamiltonian - the free Hamiltonian
 of an electron in a wire  (``external  channel") -
 is defined by the  Schr\"odinger-type
differential expression:
$$
lU =
-\frac{\hbar^2}{2m}
\frac{\mbox{d}^2U_0}{\mbox{d}^2x},
$$
where the effective electron mass $m$
is determined  both  by the wire
structure and the Fermi level $\lambda_F$
energy near the conductivity  band edge.
\par
The  quasi-molecular  cluster  attached  to  the  wire
could  be    represented of  course by
 the  equivalent  potential  well,
separated  from  the  outside  of  the  wire  by  the
potential  barriers, simulating  the  Coulomb  blockade, see [5].
But the  analysis  of   solutions  of   the  corresponding
Schr\"odinger  equation  is
generally   possible either in  the  quasi-classical  limit,
which  is rather far  from  realistic  physical  conditions,  or
for properly  choosen  soluble  models.
This  is  a  reason  why  we  use  here  a
soluble  model  based  on Operators Extensions  Theory.
\par
 We  use it here  in  form  of so  called
{\it{zero-range  potentials
with the  inner  structure}},  proposed  as  an
universal  model  for   certain class  of
Schr\"odinger-type  spectral  problems  in  [13].
These  models   supply  us  with  a  wide  range
 of  spectrally  interesting    Hamiltonians  and
 give  us  a  realistic   physical  picture
the  spectrum  and   dynamics  of  perturbed
systems, when  the
 perturbations   satisfy
the  following  condition:
\begin{quote}
 \it  the  de Broghlie  wave-length in  the
nonperturbed  system  is  greater,
than  typical  dimensions ${\delta}$ of the perturbation
considered
\end{quote}
\[
\delta<<\lambda \approx {\frac{2\pi\hbar}{\sqrt{2m kT}}},
\]
where  $m$  is  the  effective  mass of electrons
at  the   Fermi level.
\par
In a certain  sense  this condition  is  similar  to
the  condition  of the  slow  phase  change,
formulated  for  mesoscopic  processes  by  Altshuler (see [3]).
\par
Our  analysis  shows, that  the  condition  above
holds   for   constructions
of {\it  narrow-gap and zero-gap semiconductors },
 since  for  narrow-gap semiconductors
 the de Broglie wave-length,  being
 inversely proportional to $\sqrt{m}$ which
is  essentially   smaller, than  the   conventional
mass  of  electron.
\par
For the semiconductors  presented in the following Table,
the
electron wavelength is some hundreds of  Angstroms at the room
temperatures and thousands of ones at the  liquid nitrogen temperature.
Creating   relatively  ``small" clusters, which  fulfill  the  above  condition
and  can  be  considered  as perturbations  for
quantum  wires  of these  materials,  is already
 possible    using modern technical and
technological tools: electronic
lithography, ionic implantation, and selected chemical etching [14].
\par
  This   allows  our  device to
work at   essentially   higher temperatures as  known
world samples  of   quantum  gates. Our calculations (see Section 3,[15])
are
illustrated by  the following Table.
\bigskip

\begin{center}
\begin{tabular}{|l|c|c|c|c|c|c|}
\hline
Semiconductor &
\begin{tabular}{c}
\strut $E_g$ (300\,K)\\
\strut eV
\end{tabular} &
$\ds \frac{m^*_e}{m_0}$ & $\ds\frac{m^*_p}{m_0}$ &
$\varepsilon_{sc}$ & \begin{tabular}{c}\strut $\lambda$\,(300\,K)\\ \strut
\AA\end{tabular} & \begin{tabular}{c}\strut $\lambda$\,(77\,K)\\
 \strut \AA\end{tabular} \\ \hline
$GaAs$ &                 1.430  & 0.070   & 0.40 & 12.5 & 290 & \ 580 \\ \hline
$InAs$ &                 0.360  & 0.022  & 0.40 & 15.0 & 515 & 1020\\ \hline
$Cd_xHg_{1-x}Te$ & \multicolumn{6}{|l|}{} \\ \hline
$x=0.20$ & 0.150  & 0.0130 & 0.45 & 17.0 & 670 & 1320\\ \hline
$x=0.25$ & 0.220  & 0.0165 & 0.45 & 16.4 & 590 & 1160\\ \hline
$x=0.27$ & 0.250  & 0.0180 & 0.45 & 16.2 & 570 & 1125\\ \hline
$x=0.28$ & 0.260  & 0.0190 & 0.45 & 16.1 & 550 & 1085\\ \hline
$x=0.29$ & 0.275  & 0.0200 & 0.45 & 16.0 & 540 & 1065\\ \hline
$x=0.30$ & 0.290  & 0.0210 & 0.45 & 16.0 & 525 & 1035\\ \hline
$x=0.31$ & 0.300  & 0.0220 & 0.45 & 15.9 & 515 & 1015 \\ \hline
$x=0.32$ & 0.315  & 0.0230 & 0.45 & 15.8 & 500 & \ 985 \\ \hline
$x=0.44$ & 0.510  & 0.0550 & 0.45 & 15.2 & 325 & \ 640 \\ \hline
$HgTe$   & -0.117 & 0.012  & 0.50 & 21.0 & 700 & 1380\\ \hline
$Zn_{0.15}Hg_{0.85}Te$\footnotemark[1] & 0.190  & 0.015  & 0.45 & 17.0 &
620 & 1220\\ \hline
\end{tabular}
\footnotetext[1]{In series of triple compounds $A_2B_6$, this material is
more prospective in technology,
as its properties are considerably more stable in comparison with
$Cd_xHg_{1-x}Te$.}
\end{center}

\bigskip

\par
Thus the properties of the  mentioned
semiconductive materials allow
to realize the low-dimensional quantum system for temperatures
{\it higher than the nitrogen boiling temperature}. The characteristic
dimensions of such structures are available for the
theoretical  analysis  of  the  construction  of
devices  using  the  developed zero-range  potential
techniques  and  their  fabrication
by  modern technology methods.
\par
 At  the  actual  moment   the basic problem
requiring the experimental solution  remains the problem of  the
influence of charge heterogeneities near the low-dimensional
system. The possibility to achieve low densities of the charge
heterogeneity was shown in our experiments for the Ge surface
at the room temperature [18].
\par
The proposed  model  can  give  us
realistic   spectral  properties  of  objects  under
consideration, but of  course  {\it not  the  local  differential
properties  of  the  solutions}  of  the  corresponding
Schr\"odinger equation,  hence  it
 is  essentially  ``spectral  modelling".
On  the  other  hand, our  models   help  producing
a  lot  of   explicit  formulas  for  spectral
characteristics such as scattering  matrices and resonances,
both  in asymptotically  homogeneous  and  periodic  cases.
\par
According to the general  scheme of the zero-range
  potentials with the internal
structure, developed  in [13], the connection of
 the model cluster determined by matrix   $A$
and the wire is described by the boundary conditions,
 which  connect the value of
the external component of the wave function and
 its derivative jump
\[
\left[
\frac{\mbox{d}U_0}{\mbox{d}x}
\right](0)= \frac{\mbox{d}U_0}{\mbox{d}x}(+0) -\frac{\mbox{d}U_0}{\mbox{d}x}(-0)
\]
 at the point  $0$ of connection of  both  channels (the cluster and the wire):
$$
\left\{
-U_0(0),
\left[
\frac{\mbox{d}U_0}{\mbox{d}x}
\right](0)
\right\}
$$
to the  {\it  boundary values }
$\xi_\pm$ of the internal component  $U_1$ determined by the
choice of the  corresponding {\it{deficiency  vector}} $h$ in the
 {\it{internal}} (cluster)  ``channel":
$$
U_1 =
U_{10} +
\frac{A}{A-iI}\xi_+\cdot h
+
\frac{1}{A-iI}\xi_-\cdot h
$$
$$
(A-iI)U_{10}\perp h .
$$
This boundary condition is given by  some Hermitian matrix
$\{\gamma_{ik}\}$, which  is  a  parameter  of  the  model  as  well:
$$
\left(
\begin{array}{c}
-U_0(0)
\\
\xi_-
\end{array}
\right)
=
\left(
\begin{array}{cc}
\gamma_{00} & \gamma_{01}
\\
\gamma_{10} & \gamma_{11}
\end{array}
\right)
\left(
\begin{array}{c}
\left[
\frac{\mbox{d}U_0}{\mbox{d}x}
\right](0)
\\
\xi_+
\end{array}
\right).
$$
The  above  boundary  condition  describes  the
 electron's transition from the internal (cluster)   channel
to the external one   $L_2(R)$  (the  wire) and backwards.
 The operator $l_A$ which is
defined  on  the  linear  set  of   elements
$(U_0,U_1)$,  which  satisfy the  above boundary condition
$$
l_A
\left(
\begin{array}{c}
U_0
\\
U_1
\end{array}
\right)
=
\left(
\begin{array}{c}
-\frac{\hbar^2}{2m}
\frac{\mbox{d}^2U_0}{\mbox{d}x^2}
\\
AU_{10} -
\frac{1}{A-iI}\xi_+h
+
\frac{A}{A-iI}\xi_- h
\end{array}
\right)
$$
proves  to  be  self-adjoint. It  will  serve  as a Hamiltonian of
a  single   electron in a wire with a  cluster
attached  to  it. This  Hamiltonian    can be used to describe
the electron dynamics defined  by  the solutions  of  the
corresponding  nonstationary  Schr\"odinger
equation:
$$
\frac{1}{i}
\frac{\mbox{d}}{\mbox{d}t}
\left(
\begin{array}{c}
U_0
\\
U_1
\end{array}
\right)
=
l_A
\left(
\begin{array}{c}
U_0
\\
U_1
\end{array}
\right).
$$
The  constructed model is obviously soluble:
the eigenfunctions of the  Hamiltonian can be given in an explicit
form
$$
\Psi_\lambda =
\left(
\begin{array}{c}
\Psi_\lambda^0
\\
\Psi_\lambda^1
\end{array}
\right) .
$$
 In  particular, for the scattered waves we have:
$$
\Psi_\lambda^0
=
\left\{
\begin{array}{c}
\mbox{e}^{ikx} + R\mbox{e}^{-ikx},

x>0,
\\
T\mbox{e}^{ikx},

x<0;\, \lambda=k^2
\end{array}
\right.
$$

$$
R=T-1,
$$

$$
T(\lambda) =
\frac{\frac{ik\hbar^2}{m}[\gamma_{00}-|\gamma_{01}|^2(\gamma_{11}-Q(\lambda)
)^{-1}]}
{\frac{ik\hbar^2}{m}[\gamma_{00}-|\gamma_{01}|^2(\gamma_{11}-Q(\lambda))^{-1
}] - 1},
$$
\bigskip
$$
\Psi_\lambda^1 =
\frac{A+iI}{A-\lambda I}h
(Q(\lambda)-\gamma_{11})^{-1}
\frac{\gamma_{10}}{m}\hbar^2(1-T).
$$
\bigskip
Here  $Q(\lambda) \equiv
\langle
\frac{1+\lambda A}{A - \lambda I}h, h
\rangle
$
is the rational function which  has a positive imaginary
part in the upper half-plane (and the negative one in the lower
one). If the parameters  of  the  model  are connected by
the   condition
$$
\gamma_{11}
+
\langle
Ah, h
\rangle
= 0,
$$
then the  transmission coefficient $T$
has  the  physically  correct
high energy asymptotic which is typical for the Schr\"odinger
equation with smooth potentials:
$$
T(k) = 1 + o(1/k).
$$
The conductivity of the  wire  with  the  model  cluster
attached  to  it  is given by  the  Landauer
formula [10, 11]:
$$
\sigma(\lambda) =
\frac{e^2}{\hbar}
\frac{|T|^2}{1-|T|^2}.
$$
Thus the conductivity vanishes when  $T(\lambda) = 0$. In  particular,
neglecting  the  analog  of ``chemical shifts" $\gamma_{00}$,
$\gamma_{11}$  caused by the  joining  of  the
cluster  to the quantum wire    we   get:
$$
T(\lambda) =
\frac{\frac{ik\hbar^2}{m} |\gamma_{01}|^2}
     {\frac{ik\hbar^2}{m} |\gamma_{01}|^2 + Q(\lambda)},
\quad
\lambda = k^2.
$$
We  see that the conductivity of the device considered
vanishes for  electrons having the energies equal to
ones  of  the
non-occupied orbitals corresponding  energy levels
$\alpha_s$  of  the  attached  cluster. In  particular, if the
cluster is in the  ground  state,  $A=A_g$, and  the  non-occupied
energy  levels are $\alpha_e$, $\alpha_1$, $\alpha_2$, \dots,
then  the  transmission  coefficient  vanishes  for
electrons  with  the  corresponding  energies.
If it is  in the  excited  state  $A=A_e$, then  the
nonoccupied levels are equal  to           .
$\alpha_g$  $\alpha_1$ $\alpha_2$, hence
the transmission  coefficient  vanishes  for  electrons
 having these  energies.
\par
If the Fermi level $\lambda_F$ in  the  wire coincides with
$\alpha_e$,  then
in the first case the device
conductivity  vanishes
at cryogen temperature, and it is positive
in the second case:
$$
\sigma_g = 0,
$$

$$
\sigma_e =
\frac{e^2}{\hbar^2}
\frac{|T_e(\lambda_F)|^2}{|T_e(\lambda_F)|^2 - 1},
$$
where  $T_e$   is  the  transmission  coefficient corresponding
to the  excited  cluster.
\par
The  switching on  and  off of this  device  can be managed
 by   any  type  of  optical
excitations. When  excited   by  the
resonance  light, the
 cluster electron is transferred
 from the level $\alpha_g$ to the level
$\alpha_e$     and  is  supposed  to    remain  there
until  the next exposition with resonant light induces the
inverse transition.
\par
 At higher temperature the considerable part of conductivity
electrons has the energy $\lambda$ that differs from
$\lambda_F$, so the  real   conductivity  should
be  calculated  as   a
mean value of conductivity $\sigma(\lambda)$
with Fermi distribution
$\rho(\lambda,t)$   as
follows:
$$
\int\sigma(\lambda)\mbox{d}\rho(\lambda,t) =
\bar{\sigma}(t).
$$
Obviously  the  operating  of  the
described   gate   is  not
efficient  for  higher  temperatures, since  the
value of the  mean  conductivity  could  be
far  from  zero  in  ground  state.
 But  being  joined  in  a  chain,
these  gates  seem  to  be  essentially  more  efficient
even  for  higher  temperatures.
\vskip0.5cm
{\bf{3. Soluble  model for  optically   simulated  metal-dielectric
   transition  in  infinite  quasionedimensional  lattice}}.
In  this  section  we   consider the
infinite  chain  of  model  molecular  clusters,
similar  to  one  discussed  above.
We  assume, that  the clusters are chosen  such  that in the
ground state all of  them have the same Hamiltonians
$$
A^s =
A_g^s = A_g
$$
represented  by  the  finite-dimensional operators
(diagonal matrices)
in the spaces  $E^s = E$.
Further  we  assume  that
 the  optical  excitation  with  the  frequency  $\nu$
is  in  resonance    with  odd  clusters  only,
hence  only  clusters  with  the odd numbers  $s = 2l+1 $  get  excited,
\[
\alpha_e^{2l+1} - \alpha_g^{2l+1} =2\pi {\hbar}\,\,\nu,
\]
and   all  excited  clusters  are  equivalent, $A_e^{2l+1} = A_e^1$.
Thus, if in the  ground  state  the
chain  of clusters  has  the  period
$L$, then  in the excited state it  has the period  $2L$.
At  last  we  assume
that the deficiency elements $h=h_s$  and   the
self-adjoint boundary conditions describing the
connection of neighboring clusters  are
 the  same for all clusters  of  the  chain:
$$
\xi_-^s =
\gamma\xi_+^{s-1} +
{\bar \gamma}\xi_+^{s+1}.
$$
We consider  a selfadjoint operator ${\cal A}_\gamma$
defined  in  the  space ${\cal E} = l_2 (E)$ of  all  infinite
sequences $U^s$ of $E$-vectors:
$$
U = \{U^s\}_{s=-\infty}^{s=\infty}\subset
{\cal E}=\sum\oplus E_s .
$$
Writing  down  each  component in a  form
parametrised  by  the  boundary
values, as  it   was  done
in  the  previous  section
$$
U^s =
U_0^S
+
\frac{A^s}{A^s - iI}\,\,\xi_+^sh
+
\frac{1}{A^s - iI}\,\,\xi_-^s h
$$
we  define  the  operator by the formula
$$
({\cal A}_\gamma U)^{(s)} =
A^sU_0^s -
\frac{1}{A^s - iI}\,\,
\xi_+^s h
+
\frac{A^s}{A^s - iI}\,\,\xi_-^s h
$$
and the  proper translation  invariant
self-adjoint  boundary conditions
imposed  onto  the  boundary  values  $\xi_{\pm}^s$:
\[
\gamma \xi_{+}^{s-1} +{\bar\gamma}\xi_{+}^{s+1} = \xi_{-}^s.
\]
Then  the operator  ${\cal A}_\gamma^g$ is the
one-particle  lattice model with the
 period  $L$, and the
operator ${\cal A}_\gamma^e$  is the lattice model with the period $2L$.
 In  the generic case the spectrum of the operator  ${\cal A}_\gamma^g$
consists
of the  dim$A_g$   bands which are determined by the
solutions  of  the  algebraic  equation:
$$
Q_g(\lambda) =
\langle
\frac{I +\lambda A_g}{A_g - \lambda I}h,h
\rangle
=
\gamma\theta^{-1} - {\bar \gamma}\theta,
$$
with the uni-modular unknown  function $\theta = \mbox{e}^{i\kappa}$,
$\kappa$  playing  the  role  of  quasimomentum  exponential  $e^{i\kappa}$,
  and
\[
U = \left\{U^s \right\}= \left\{U^0 \theta^s\right\}
\]
generally  playing  the  role  of  Bloch-type  solutions  of
the  corresponding  homogeneous  equation  and
the  role  of  eigenfunctions   of ${\cal A}_{\gamma}^g$
on  the  corresponding  spectral  bands.
The condition of  existence  of  nontrivial bounded
solutions  of  the  last   algebraic  equation takes the form
$$
|Q_g|\equiv
\left|
\langle
\frac{I +\lambda A_g}{A_g - \lambda I}h,h
\rangle
\right|
\leq
2 |\gamma| .
$$
The  spectrum  of  the  operator  ${\cal A}_{\gamma}^g$  consists   of
spectral
bands, constituted  by the (real)  values  of  the  energy  $\lambda$,
which  satisfy  the  last  condition.
Two Bloch eigenfunctions correspond to each value $\lambda$  from the band:
$$
\Psi_\pm^s =
\frac{A^g + iI}{A^g -\lambda I}
L\theta^{\pm s},
\quad
s=0,t,\dots .
$$
According  to the  conventional
 quantum-mechanical  interpretation the  electrons
 in the Bloch  states, corresponding
to  spectral  bands  of  nonexcited  chain
  are   not  ``localized"  but
are  ``moving" in  the
infinite  one-dimensional conductor formed by the periodic
chain of clusters in the ground state. Under the values  $\lambda$,
lying outside of the spectrum $|\theta(\lambda)|<1$
or $|\theta(\lambda)|>1$, the  corresponding  Bloch
solutions $\Psi_\pm$
are  exponentially  decreasing   in one  direction and exponentially
growing in  the  opposite  one so  they  can't be  eigenfunctions
of   the  continuous  spectrum. In  real  lattices
 the  electrons with the corresponding energies  are either  absent
 or  localized on impurities in  the  proper  bound  states,which  are
combined  of  the   decreasing  branches  of  the  corresponding
Bloch  solutions.
\par
If each odd cluster is  excited, and each even
one is in the ground state
$$
A^{2l} = A_g,
$$
$$
A^{2l+1} = A_e,
$$
then  the corresponding Hamiltonian ${\cal A}_\gamma^e$
is also self-adjoint one, and its
spectrum  also has  a band  structure. However, the condition of belonging
$\lambda$ to the spectrum of the $2L$-periodic lattice is described
technically  in
other  form. Now  the  role  of  the  period  of  the unperturbed
 lattice  operator
is  played  by  the  orthogonal  sum  $A_g \oplus A_e$  in  $E\oplus E$,
and  the  Bloch-vector  of  the  perturbed  lattice  with  the  boundary
conditions  switched  on  is  represented  as  a  sequence  of
two-component  $E$-vectors
$$
\Psi =
\{
\theta^n
(
\begin{array}{c}
v_0
\\
v_1
\end{array}
)
\}.
$$
 Denoting  by  $A_{e,g}^*$  the
formal  expression  for  the  operator's  action  on  elements
of  $E$, represented  as  in  the  first  section,
$$
U_1 =
U_{10} +
\frac{A}{A-iI}\xi_+\cdot h
+
\frac{1}{A-iI}\xi_-\cdot h,
$$
$$
(A-iI)U_{10}\perp h,
$$
we  write down  the  corresponding  homogeneous  equation
for  the  boundary  values $\xi_{+}, \xi_{-}= Q(\lambda)\xi_{+}$
of  Bloch  vectors of  the   exited   chain
\[
\Psi_{\pm} = \left\{
\begin{array}{c}
\frac{A_g +iI}{A_g - \lambda I}\xi_{+}^g\\
\frac{A_e +iI}{A_e - \lambda I}\xi_{+}^e
\end{array}
\theta^{\pm l}
\right\}_{l= -\infty}^{l= +\infty}.
\]
 in  the  following form
\[
\left[
\begin{array}{cc}
Q_g (\lambda) &-\bar\gamma - \gamma\bar\theta\\
-\gamma - {\bar\gamma} \theta & Q_e (\lambda)
\end{array}
\right]
\left(
\begin{array}{c}
\xi_g\\
\xi_e
\end{array}
\right)_{+}
= 0,
\]
\[
\left[
\begin{array}{cc}
Q_g (\lambda) &-\bar\gamma - \gamma\theta\\
-\gamma - {\bar\gamma}\bar\theta & Q_e (\lambda)
\end{array}
\right]
\left(
\begin{array}{c}
\xi_g\\
\xi_e
\end{array}
\right)_{-}
= 0.
\]
In  particular, from  the  second  equation  we  have  the  following
condition  for  the corresponding
 quasi-momentum  exponential $\theta = e^{i\kappa}$
\[
Q_g (\lambda) Q_e (\lambda) = 4 |\gamma |^2  cos^2 (\varphi +
\frac{\kappa}{2}).
\]
Here  $\varphi = {\mbox {arg}}\gamma$. Thus $cos^2 (\varphi + \frac{\kappa}{2})=
\pm \frac{\sqrt {Q_g Q_e}}{2 |\gamma|}$, and
\[
\xi_g^- = \frac{1}{\sqrt {|Q_g|}},\,\,\xi_e^- = \frac{e^{i\kappa}}{\sqrt
{|Q_e|}}.
\]
It  follows  from  here, that  the  spectral  bands
of  the  excited  lattice  coincide  with
the  intervals  of  real  axis, where
 $0\leq Q_g (\lambda) Q_e (\lambda) \leq 4 |\gamma |^2 $.
Another  important  fact  concerning  the  boundary  values $\xi_{g,e}^e$
of   Bloch   vectors  will  be  used  in  the
 following  section  for  estimates  of  the
transmission  coefficient:
\par
   {\it The linear  combination of  the  boundary  values  with  pure  imaginary
   coefficient  $k$:
\[
   \xi_g^-  + ik \gamma \theta \xi_e^-
\]
 does  not  have  roots  on  the  real  axis  $\lambda $.}
\par
Assuming that the spectral properties of clusters in the ground
and excited states  are the same as in the previous section  we see
that the functions $Q_g$, $Q_e$
are essentially distinguished by the position
 of  one   pole   only:
$$
\frac{1+\alpha_e^2}{\alpha_e - \lambda}
|\langle
h,\phi^e
\rangle|^2
\longleftrightarrow
\frac{1+\alpha_g^2}{\alpha_g - \lambda}
|\langle
h,\phi^g
\rangle|^2 .
$$
This leads to the partition of  the  lower  spectral  band
 of the $L$-periodic chain of clusters
 $A_g$ into  two subbands  corresponding to the $2L$-periodic
 chain  of $A_g,A_e$-clusters. The arising
lacuna  (spectral gap)  $\delta$ is found from the condition:
$$
Q_e(\lambda)Q_g(\lambda)<0,
$$
and shifts $\lambda_e^\pm\to\lambda^\pm$
of the upper and lower bounds are determined from the
equation:
$$
Q_e(\lambda)Q_g(\lambda)=4|\gamma|^2.
$$
If the residues $(1+\alpha^2)|\langle h,\phi^e\rangle|^2$
at the poles $\alpha_e$ $\alpha_g$,  which  correspond  to the ground
and excited states of the clusters, are approximately equal,
then  the  dispersion  curves $\lambda = \lambda (\kappa)$ corresponding  to
the  excited  and the nonexcited chain  are
approximately  parallel  and   hence  the lacuna (spectral  gap)
width  $|\delta|$ proves to be approximately equal to the distance between the
ground  and  exited  energy levels  of  odd   clusters:
$$
\delta\sim\alpha_e - \alpha_g .
$$
\par
The  described  chain  exhibits  a  remarkable  type  of
  behavior  under  resonance  optical  excitations, which  can  be
interpreted  as  a   ``{\it{Simulated  Mott-Pejerls  transition}}".
 If the Fermi level $\lambda_F$  is  situated at
the  center of  the  gap $\delta$ (thus {\it inside} the  conductivity
 band of the chain  in  ground  state ),
then  the   chain in the ground  state is  equivalent  to  one-dimensional
metallic  conductor. On  the  other  hand  the  Fermi
level  is  situated  inside  the  gap  $\delta$  of  the
 exited  chain, hence  the  {\it{exited  chain  is  a
dielectric  one  for  electrons  on  Fermi  level $\lambda_f$}}.
This  is  exactly  Mott-Pejerls-type  behavior, see [19].
It  is  important, that the  chain  remains
dielectric  within  some  interval  of  energies  near $\lambda_f$,
defined  by  the  width  of  the  gap $|\delta|$,
hence  the  described  chain  {\it{can  plays  the  role
of  the  gate  in  some  interval of  temperatures}}
 $0\leq T\leq \frac{|\delta|}{2k} $. Under  conditions  above this interval
 is  approximately  determined by the distance between the levels
$\alpha_g - \alpha_e$.
In  section  2  the  advantages  of  narrow-band  semiconductors
were  discussed  for  creating  quantum  wires  with  large
values  of  de  Broghlie  wave length. Here  we  underline
another essential advantage of the the narrow-gap
semiconductors with small values $m$ :  they   manifest  a high
resolution of the levels in quantum well which can exceed $kT$ already
in the range of room temperatures. It  means, that  the
distance  between  energy-levels  of  cluster  orbitals
is  large  , and  thus, according  to  observation  above,
the  critical  temperature   $\frac{|\delta|}{2k}$
 may be significantly higher than the helium
temperature  or  even  higher, then the  nitrogen  one.
\vskip0.5cm
{\bf 4. Scattering  by finite  chain  of  clusters.}
The  quantum  gate  can't  be
technologically implemented  in the form  of
infinite   cluster  chain.  We  consider  now  the  finite
  chain  of  clusters  length $N\,\,L = 2n\,\,L$ inserted  into  a  quantum
wire.
Assuming  that  the  clusters  of  the  chain  possess properties
described  in  the  previous  section, we solve  the
 scattering   problem for  exited  chain  and  for the
chain  in  ground  state. The corresponding transmission
coefficient  proves  to  be  exponentially
small  for  large  $N$ outside  the  spectral  bands  of
the  exited infinite  chain, but  is  close  to
the  transmission  coefficient  ``through  the  semi-infinite  chain"
in the ground  state.
 Practically  it  means  that  the  finite  chain  inserted
 in  the  quantum  wire  can  serve  as  a  quantum  gate
under  certain  conditions on  the  parameters  of  the
clusters  and  the  chain.
\par
We  construct the model Hamiltonian  of  a  single  electron on  the
quantum  wire  with  the  inserted  finite  lattice
 by  means  of  extension  theory  methods
of two  components: the  free  Hamiltonian  on  the  wire  and  lattice
Hamiltonian  on  the  inserted  periodic  chain  with
proper  boundary  conditions  at  the  points  of  contacts:
\[
\left(
\begin{array}{cccccccccc}
0   &\bar\gamma & 0 & 0 & ... & ... & ... &  0  &   0  &  0\\
\gamma & 0 &\bar\gamma &  0  &  0  &...&...&... & 0 & 0  \\
 0  & \gamma  &  0  & \bar\gamma & 0 ...&...&...&...& 0 & 0\\
 ...&...&...&...&...&...&...&...&...&...\\
 ...&...&...&...&...&...&...&...&...&...\\
 ...&...&...&...&...&...&...&...&...&...\\
 ...&...&...&...&...&...&...&...&...&...\\

..&...&...&...&...&\gamma & 0 & \bar\gamma & 0  & 0 \\
..&...&...&...&...&...&\gamma &  0  &\bar\gamma & 0\\
..&...&...&...&...&...&...& \gamma & 0  &\bar\gamma \\
..&...&...&...&...&...&...& ...& \gamma  &  0
\end{array}
\right)
\left(
\begin{array}{c}
u'(0)\\
\xi_g^0\\
\xi_e^0\\
\xi_g^1\\
..\\
..\\
\xi_e^{n-1}\\
\xi_g^{n}\\
\xi_e^{n}\\
-u' (N L)
\end{array}
\right)
=
\left(
\begin{array}{c}
u(0)\\
Q_g\xi_g^0\\
Q_e\xi_e^0\\
Q_g\xi_g^1\\
..\\
..\\
Q_e\xi_e^{n-1}\\
Q_g\xi_g^{n}\\
Q_e\xi_e^{n}\\
u (N L)
\end{array}
\right).
\]
The  scattered  waves are  combined   of  the  solutions  of
the  stationary  Schr\"odinger  equation  on  the  quantum  wire
$$
\begin{array}{cc}
T e^{ikx},\,\,& -\infty < x \leq 0,\,\,\\
e^{ikx} + R e^{-ikx},\,\,& N\,L\leq x <\infty,
\end{array}
$$
 and  Bloch  solutions
on  the  inserted  lattice $ 0 \leq l \leq N $:
\[
\Psi = \alpha \Psi_+  +  \beta \Psi_-.
\]
A  straightforward  calculation  gives  us  the
connection  between  $\alpha$  and  $\beta $:
\[
\frac{\beta}{\alpha}=
-\frac {\theta^{-1}\xi_e^+ - ik\gamma \xi_g^+}{\theta \xi+g^- -ik\gamma\xi_g^-},
\]
and   the   explicit expression  for  the  transmission  coefficient.
 Let  us  denote
\[
\Delta_+,\,\, \equiv
\frac{\theta\xi_g^{+} + ik\gamma \xi_e^{+}}
{\theta^{-1}\xi_e^{+} - ik\gamma\xi_g^{+}},\,\,\,\,
\Delta_- \equiv
\frac{\theta^{-1}\xi_g^{-} + ik\gamma \xi_e^{-}}
{\theta\xi_e^{-} - ik\gamma\xi_g^{-}}.
\]
Then
$$
T = \frac{-2ik\gamma e^{iL N k}}{\theta \xi_e^{-} - ik\gamma e_g^-}
\frac{\xi_g^{+}\xi_e^{-}\theta - \xi_g^{-}\xi_e^{+}\theta^{-1}}
{\theta^n \Delta_+ - \theta^{-n} \Delta_-}.
$$
Due  to  the  statement  from the  previous  section  the  coefficient  in
front  of  $\theta _n$  in  the  denominator  does  not  have  real  zeroes,
hence  the  transmission  coefficient  has  the  following  asymptotic
on  spectral  gaps  of the exited chain  for  large  $N$:
\[
T \approx 2ik \frac{\bar\gamma}{\gamma}
\frac{e^{ikL N}\theta^n}{\theta^{-1}\xi_g^- +ik\gamma \xi_g^-},
\]
so  it  is  exponentially  small  inside  the
  spectral  gaps  of  the  exited  chain.
\par
On  the  other  hand, the  similar  calculation  of  the  reflection
coefficient $R^N$ of  the  chain  in  the  ground  state  gives
\[
R^N_g = e^{2ikL N}\frac{\theta^N \Delta_1 - \theta^{-N} \Delta_2}
{\theta^N \Delta_3 - \theta^{-N}\Delta_4},
\]
where
\begin{eqnarray*}
&\Delta_1 \equiv \frac{Q + ik|\gamma|^2 - \theta^{-1}\gamma }
{Q - ik|\gamma |^2 - \theta \bar \gamma},\\
&\Delta_2\equiv\frac{Q + ik|\gamma|^2 - \theta \gamma }
{Q - ik|\gamma |^2 - \theta^{-1} \bar \gamma},\\
&\Delta_3 \equiv\frac{Q - ik|\gamma|^2 - \theta^{-1} \gamma}
{Q - ik|\gamma |^2 - \theta \bar \gamma},\\
&\Delta_4\equiv\frac{Q - ik|\gamma|^2 - \theta \gamma }
{Q - ik|\gamma |^2 - \theta^{-1}\bar\gamma}.
\end{eqnarray*}

Note that  $|\Delta_{1,2}|=1$, hence  the  reflection  coefficient
corresponding  to the  inserted  into  quantum  wire
finite  chain  of $ N $ equivalent  non-excited  clusters
  has  exactly  $ N $  roots
on  each   spectral  band  of  the  corresponding  infinite  lattice. Due  to
the  unitarity  of  the  complete  scattering  matrix  it  means that  the
corresponding  transition  coefficient  $T_g$ is modulo  one  at  $N$  spectral
points  on  each  spectral  band  of  the  corresponding  infinite  lattice.
\par
The  Reflection  Coefficient $R^{N}_g (k)$ is  a  contracting  analytic
function
on  the lower halfplane  of  momentum $k$. It  can  be  shown,
that  on  every  compact  domain  of  it  it  is  converging  uniformly  to  the
reflection  coefficient $R_g^{\infty}$, corresponding  to  the
semi-infinite  lattice.
>From  the  observation  above  follows, that  this  convergence  does  not
take
place  on  real  axis  of  momentum,in  particular
  on  spectral  bands  in  the  complex
plane   $\lambda$  of  energy. Nevertheless  for  any   problem
for  the  corresponding  nonstationary  Schr\"odinger  equation  with   smooth
initial  data  given the in  form  of  incoming  waves  in  quantum  wire
one  can
use  the  fact, that  both  reflection  coefficients  are  close  ``in  medium"
for  large  $N$:
\[
R^N_g(k) \approx R_g^{\infty}.
\]
Thus  the  measure  of  ``formal  garbage"  produced  by
   any  scattering   act
by  finite  chain  is  limited  by
the  corresponding  reflection  coefficient  of  the  semi-infinite
lattice. But  similarly  to  the  the  situation  described  in  the  section  2
we  see that  the  information  on  the  reflected  waves  is  contained
actually   in  the  transmitted  waves, so  it  is  never  lost,
it  is  `` potentially  observable".
\par
The  real  garbage  in  a  single  act  of  computation  is  caused  by
the  measurement  of  the  scattering  state, i.e. by  projection  of
the  state  onto  the  states  of  measuring  device, or  by  averaging
them  over  stochastic  states  of  the  random  background. The  loss
of  this  garbage  is  unrecoverable. The  corresponding  information
is  irreversibly  lost  which   makes  the  computation
irreversible.
  In  [18]    efficient  probabilistic  algorithms  for  classical  computations
are  discussed. Similarly  to  the  discussed  quantum  computations, these
algorithms  are  ``polynomial-time  algorithms". Professor C. Calude
formulated  the following question: Are there  any  `typical
features`  of  quantum  computations which make them different,
and more efficient, than     classical  probabilistic computations?
 Our  model permits  to  compare  the work of the  described  quantum
  gates  with  the  corresponding  classical  gates  as  an  element
of  classical  circuit  acting  in  frames of a ``random"
algorithm. Really, the  single  act of  quantum
computation  consists  in  opening  or  closing   the quantum  gate.
Assuming that  the  life-time   of     inner  cluster  electrons  on
the  excited  level  is   much  longer than  typical  time-intervals
of the  process  under  consideration, we  see  that  the  the  quantum
evolution  of  the  conductivity  electron  is   unitary,  and  the
computation  is  invertible, unless  we  do  not   project  the  result
onto  the  states  of  the  measuring  device. {\it The  inversibility   of
the  computation  is  produced  by  the  act  of  measurement  only}.
  \par
If  the  constructed  self-adjoint  operator is  positive, we  can  consider
it  as  a  generator  of  evolution  of  the  transition  probabilities
of  some  Markov  process. Then the  relevant  parabolic  equation
is  interpreted  as  a   Fokker-Plank  equation for  transition
probabilities. The  corresponding  evolution  is
contracting, hence  the  conservation  law  is
absent, and {\it the  garbage, produced by  the  process, is  killed  each
moment  in  the course  of  evolution, resulting  in  the  global
irreversibility  of  the  process}.
So the important difference  between  the  quantum  computations
and  similar  probabilistic  algorithms  is  the  moment  when  the
unrecoverable  garbage  is  produced, which  corresponds  to
the  {\it irreversible}  loss  of  information. For  classical  computations
it  is  produced  {\it in  course  of  evolution}, and  for  quantum
ones it  is  produced just  in  the  last  moment. The
``potentially  observable  garbage",
caused  by  outgoing  waves in  quantum   process does  not   destroy
 the  reversibility
of  the  process  of  quantum  computations, though  these
 waves  remove  some  part   of  wave-packet  to  infinity.
Possibly this  difference  can  help  defining the  proper
domain  of  applications  of  both  computational  techniques.
\par
The implementation of the
 construction  of  quantum   gates
described in  section  4   and  other  constructions  based  on
 simulated Mott-Pejerls
transitions (see [20])
 require  solving  numerous
 difficult physical and technological
 problems such as the choice
of proper quasi-molecular clusters and their
 location inside the isolating
medium, the information output/input from the macroscopic level, the
choice  of the  proper
optical window for each type of gates etc. Possibly,
the most prospective constructions of such kind may be created as
polymers or biomolecules. The general advantage of the Mott-Pejerls
gates is  higher  stability  of  their
characteristics and a  relatively  weak
 dependence of   their working  regimes on
temperature.
\par
{\bf Acknowledgment.}
The  authors  are  grateful  to   Professor  P. Zoller  for
the  materials  supplied and  to   Professor  K. Svozil   for
the  fruitful  discussion.
\par
The  authors are grateful  to  Professor  C. Calude who
read  and  commented  the first  variant  of  the text  and  formulated  the
interesting  question.
\par
This  paper  was  written
as a part  of the  Collaboration  Project managed
by  Solvay  Institute,  under support  of the  system
of Grants  ESPRIT,  awarded  by  the  Commission  in  Research  and
Development  of  European  Community.

\newpage

\begin{center}
{\large\bf      References}
\end{center}
\bigskip
\noindent
[1] P.Shor,  ``Algorithms for Quantum Computation: Discrete Logarithms and
     Factoring".  In {\it{ Proceedings 35  Annual Symposium  of  Computer
     Science,}} edited  by  S. Goldwasser (IEEE  Computer  Society  Press,
     Los  Alamitos,  CA, 1994), p.124.
\vskip0.5em
\noindent
[2] R.P. Feynman,  Quantum  Mechanical  Computers. {\it Optics News},  {\bf 11}
     (1985), p.11.
\vskip0.5em
\noindent
[3] B.L.Altshuler, P.A.Lee and  R.A.Webb, edds.,   {\it Mesoscopic Phenomena
    in  Solids}, North Holland, New-York (1991).
\vskip0.5em
\noindent
[4] T.Ando,   A.B. Fowler, and  F.Stern,                         Rew. Mod. Phys.
    {\bf 54} (1982),p.437.
\vskip0.5em
\noindent
[5] H.Grabert, M.H.Devoret,  {\it Single Charge Tunneling - Coulomb Blocade
    Phenomena in Nanostructures}. Plenum Press, New  York (1992).
\vskip0.5em
\noindent
[6] M. Tinkham,    {\em Introduction to Superconductivity}.  McGraw Hill,
New York
    (1995).
\vskip0.5em
\noindent
[7] J.I.Cirac,  P.Zoller, Quantum  computations with  cold  trapped  ions.
{\it  Phys. Rew. Letters}, {\bf 24} (1995) p.4091.
\vskip0.5em
\noindent
[8] P.Berman, ed.    {\it Advances in Atomic,Molecular and Optical  Physics}
    Suppl.2. Academic  Press, New York (1991).
\vskip0.5em
\noindent
[9][ S. Ishihara,  {\it Optical Computers. New Science Age.} Ivanami Shoten
     Publishers. 1989.
\vskip0.5em
\noindent
[10]. R. Landauer, {\it  IBM Jornal for Research and Development} 1 (1957),
1. C.223.
\vskip0.5em
\noindent
[11]. M.Buttiker,  Nanostructured systems, ed. M.A.Reed, NY Academ, 1990.
\vskip0.5em
\noindent
[12]. Yu.Melnikov, B.Pavlov,  Two-body scattering on a graph and application
     to simple nanoelectronic device.{\it  J. Math. Phys.}  1995. Vol.36. (6),
     P.2813-2825.
\vskip0.5em
\noindent
[13]. B.S. Pavlov, The theory of the extensions  and exactly soluble models,
UMN, 1987.v.42.
\vskip0.5em
\noindent
[14]. A.M.Yafyasov, V.B.Bogevolnov, T.V.Rudakova,
    {\it Quantum interferentional
     electronic transistor (QIET). Physical principles and technological
     limitations.} Draft. St. Petersburg University (periodical report,1995).
[15] A.M. Yafyasov, V.B. Bogevolnov, T.V. Rudakova,  {\it Quantum
interferentional
electronic transistor (QIET). Theoretical analysis of electronic properties
for low-dimensional systems } (Progress report, 1995) Preprint IPRT N 99-95.
\vskip0.5em
\noindent
[16] A.M. Yafyasov, V.B. Bogevolnov, T.V. Rudakova,  {\it Low-dimensional
effects
in the Ge semiconductor surface at room temperature.} Thesis of ICSOS-5.
July 8-12, 1996. Aix en Provence, France.
\vskip0.5em
\noindent
[17] P.P.Konorov, A.M.Yafyasov, V.B.Bogevolnov,  {\it Low-dimensional effects
    on the GaSb Semiconductor Surface.} Phys.Low-Dim.Structures. 2/3.
    1995. P.133-138.
\vskip0.5em
\noindent
[18] C. Calude, {\em Information and  Randomness}. Springer-Verlag, 1994.
\vskip0.5em
\noindent
[19]. R.E. Pejerls,   {\it  Quantum Theory of Solids}. London, 1955.
\vskip0.5em
\noindent
[20]. B.S.Pavlov, G.P. Miroshnichenko,
Patent Application  5032981/25  (0103431) (Russia) from 12.03.1992.
\end{document}